# Low-temperature structural study of smectic $C_A^*$ glass by X-ray diffraction

Aleksandra Deptuch[1,*], Marcin Kozieł[2], Marcin Piwowarczyk[1], Magdalena Urbańska[3], Ewa Juszyńska-Gałązka[1,4]

[1] Institute of Nuclear Physics Polish Academy of Sciences, Radzikowskiego 152, PL-31342 Kraków, Poland

[2] Faculty of Chemistry, Jagiellonian University, Gronostajowa 2, PL-30387, Kraków, Poland

[3] Institute of Chemistry, Military University of Technology, Kaliskiego 2, PL-00908 Warsaw, Poland

[4] Research Center for Thermal and Entropic Science, Graduate School of Science, Osaka University, 560-0043 Osaka, Japan

* corresponding author, aleksandra.deptuch@ifj.edu.pl

**Abstract**

The liquid crystalline compound, forming the glass of the smectic $C_A^*$ phase, is investigated by the X-ray diffraction in the 18-298 K range. The characteristic distances within the smectic $C_A^*$ phase are determined. The electron density profile along the smectic layer normal is inferred and compared with the results of the density functional theory calculations. Observations of the selective reflection of the visible light investigate the helical ordering within the smectic $C_A^*$ glass. The results indicate slow evolution of the smectic layer spacing, intermolecular distances, and electron density distribution below the glass transition temperature. Meanwhile, the relative range of the short-range order within the smectic layers and the helix pitch are relatively constant in the glassy state.

## 1. Introduction

The smectic phases are the type of thermotropic liquid crystalline phases characterized by the lamellar positional order of molecules [1-4]. In some smectic phases, the switching in the electric field (ferro-, ferri-, or antiferroelectricity) is observed. Therefore, the smectogenic compounds are investigated for their potential application in liquid crystal displays [5-9]. The fluorinated liquid crystalline compounds abbreviated as $3FmX_1PhX_2n$ (Figure 1) were also synthesized initially for this purpose. The $3FmX_1PhX_2n$ compounds form the smectic $C^*$ phase with ferroelectric properties and the smectic $C_A^*$ phase with antiferroelectric properties in the surface-stabilized geometry; many of them exhibit both phases: $SmC^*$ at higher temperatures and $SmC_A^*$ at lower temperatures [10-14]. The $SmC^*$ and $SmC_A^*$ phases show, respectively, the locally synclinic and anticlinic order of the tilt angle of chiral molecules within layers [5]. On the scale of hundreds of nanometers, the azimuth of the molecular tilt changes helically, and the homeotropically oriented sample exhibits selective reflection in the ultraviolet, visible, or near-infrared range [10,13]. The helix axis is perpendicular to the smectic layers. Consequently, the effect related to the changes in the helix pitch is visible for a sample in the homeotropic alignment, with the smectic layers parallel to the sample's plane [5].

Later studies prove that some $3FmX_1PhX_2n$ compounds show very good glassforming properties [14-16], which makes them suitable compounds for the basis studies of the glass transition in partially ordered phases. Although there are many reports of the smectogenic small-molecule (non-polymeric) glassformers, to give as examples [14-23], the structural studies deep in the glassy smectic state are scarcer [14,24-28]. Based on a literature review, the lowest reported temperature for X-ray diffraction data on the smectic glass is 83–85 K [26-28].

The 3F5HPhF6 compound, investigated herein, has a low melting temperature of the crystal phase, equal to 301.3 K [10]. It is a good glassformer, where the crystallization on cooling is not observed, and the glass of the SmC$_A$* phase is formed [14]. This study presents the X-ray diffraction data obtained for the supercooled and vitrified SmC$_A$* phase of 3F5HPhF6 down to 18 K. The smectic layer spacing, intermolecular distance, and correlation length of the intra-layer order are determined. Three variants of the distribution of the electron density along the smectic layer normal are inferred from the integrated intensities of the diffraction peaks from the smectic layers and compared with the electron density distribution estimated from the results of the density functional theory calculations. Additionally, the selective reflection of the visible light is investigated down to 173 K to determine if the helical pitch changes below the glass transition temperature.

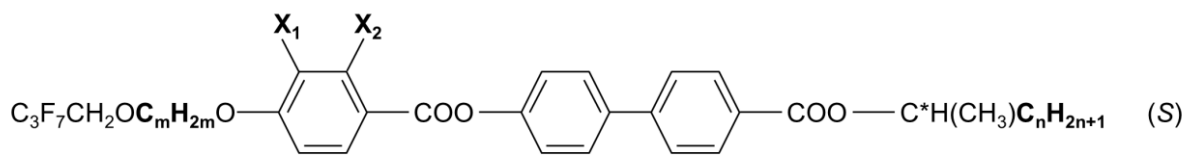

Figure 1. The general molecular formula of 3FmX$_1$PhX$_2$n. In this work, m = 5, X$_1$ = H, X$_2$ = F, n = 6.

## 2. Experimental details

The (S)-4′-(1-methylheptyloxycarbonyl)biphenyl-4-yl 4-[5-(2,2,3,3,4,4,4-heptafluorobutoxy) pentyl-1-oxy]-2-fluorobenzoate, denoted as 3F5HPhF6, was synthesized according to the route presented in [10].

The differential scanning calorimetry (DSC) measurements were performed using the TA Instruments DSC 2500 calorimeter for a sample weighing 7.17 mg in a hermetic aluminum pan. The thermograms were registered during cooling and heating at the 5, 10, and 20 K/min rates between 173 K and 403 K. The DSC data analysis was done in TRIOS.

The polarizing optical microscopy (POM) measurements in the transmission mode were carried out using the Leica DM2700 P microscope with the Linkam temperature attachment for a sample placed between two thin glass slides without the aligning layer. The images were taken every 6 s during cooling and heating at 10 K/min in the 188-393 K range. The selective reflection was investigated in the reflection mode using the same equipment for a sample placed within the electro-optic cell WAT-3A with a thickness of 5 μm and the polymer layer providing the homeotropic alignment. The images were taken every 5 s during cooling and subsequent heating at 10 K/min in the 173-377 K range and 1 K/min in the 367-377 K range. The weighted average luminance of textures was calculated in TOApy [29], and the average contributions of the red, green, and blue components were calculated in ImageJ [30].

The X-ray diffraction (XRD) measurements were performed using the Malvern Panalytical Empyrean 3 diffractometer equipped with the Oxford Cryosystems PheniX cryostat (Bragg-Brentano geometry, CuKα radiation, $\lambda$ = 1.540562 Å [31], measurement over the range of 2θ = 2-40°). The sample was initially heated to 313 K to melt the crystal phase. Then, the XRD patterns were registered on cooling in selected temperatures from 298 K to 18 K and upon subsequent heating to 298 K. The XRD data analysis was done in WinPLOTR [32] and OriginPro.

The density functional theory (DFT) calculations were done in Gaussian 16 [33] for an isolated 3F5HPhF6 molecule with the B3LYP-D3(BJ) exchange-correlation functional [34-37] and def2TZVPP basis set [38]. The starting model was prepared in Avogadro [39], and the optimized model was visualized in VESTA [40].

## 3. Results and discussion
### *3.1. Differential scanning calorimetry*

The DSC thermograms of 3F5HPhF6 (Figure 2) show a peak with the onset temperature [41] at 372.2 K on cooling and 371.4 K on heating and the average enthalpy change of 6.2 kJ/mol. This peak corresponds to the Iso/SmA*/SmC*/SmC$_A$* transitions, where Iso is the isotropic liquid. Only at the 5 K/min rate, the minor anomaly corresponding to the SmC*/SmC$_A$* transition is visible, with the onset temperature of 369.3 K on cooling and 370.8 K on heating and the average enthalpy change less than 0.1 kJ/mol (inset in Figure 2). The earlier XRD results at high temperatures [42] show that the SmA* phase of 3F5HPhF6, present only in a very narrow temperature range, is the de Vries phase, i.e., the SmA*/SmC* transition occurs with a negligible change in the smectic layer spacing. Such a continuous transition is not visible in the obtained DSC thermogram. The glass transition temperature of the SmC$_A$* phase, determined at the half-height of the step in the heat capacity [43], is equal to 227 K on cooling and 231 K on heating, with the average step in the heat capacity of 0.17 kJ/(mol·K). Noteworthy, in the earlier DSC results reported in [14], the step in the heat capacity was not visible on cooling due to a bent baseline, and $T_g$ was estimated from the kink in the DSC thermogram, leading to lower values. The $T_g$ values obtained in this study at the half-height of the step in the heat flow are in better agreement with $T_g$ = 229-230 K determined from the α-relaxation time in the dielectric spectra of 3F5HPhF6 [15].

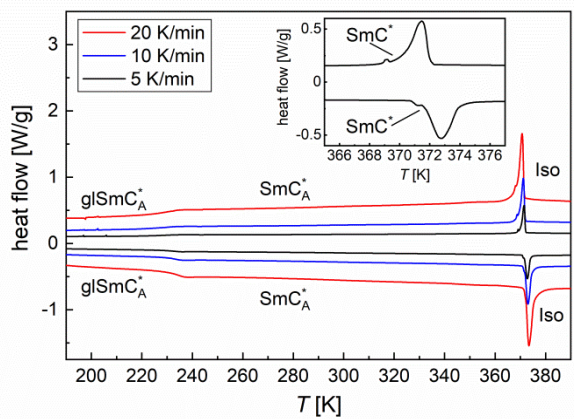

Figure 1. DSC thermograms of 3F5HPhF6. The inset shows the close-up of the SmC*/SmC$_A$* transition for the 5 K/min rate.

### 3.2. Polarizing optical microscopy and selective reflection of visible light

The representative POM textures of 3F5HPhF6, registered in the transmission mode, and the corresponding luminance vs. temperature plots are shown in Figures S1-S4 in Supplementary Materials (SM). The SmC* phase shows a green texture, while POM does not detect the SmA* phase. Despite the absence of the aligning layer, the alignment of the sample is mainly homeotropic because the helix inversion (the increase of the pitch, unwinding of the helix, and subsequent decrease of the pitch with temperature) within the SmC$_A$* phase [10] is observed as a significant change in luminance. The helix inversion temperature shows some hysteresis, occurring at ca. 325 K on cooling and 337 K on heating. The glass transition leads to a minor, step-like increase in luminance. The $T_g$ values estimated from the POM observations are equal to ca. 240 K on cooling and 245 K on heating, higher than those obtained by DSC. In the heating run, the disordered texture arises above $T_g$, see the texture registered at 290 K in Figure S3. The DSC results do not indicate the cold crystallization. Thus, such POM observation is explained by the nucleation, which is not followed by the crystal growth. The appearance of very small crystallites can disturb the alignment in the SmC$_A$* phase, but it will not be visible in the DSC scan. The texture returns to the homeotropic alignment at ca. 310 K, about 10 K above the melting temperature of the crystal phase reported in [10].

The images obtained in the reflection mode on cooling at 10 K/min and the numerical analysis results are presented in Figures 3 and 4. The corresponding results for heating at 10 K/min are shown in Figures S5 and S6 in SM. The reflection of the green light is observed for the SmC* phase (for more detailed observations within the SmC* phase, performed at the 1 K/min cooling/heating rate, see Figures S7 and S8 in SM). In the SmC$_A$* phase, the dark texture is observed. The helix inversion at 314 K on cooling and 345 K on heating shows a peak in the luminance. The hysteresis in the helix inversion temperature is wider than for the sample between glass slides without aligning layers. The helix pitch values presented in [10] down to 283 K imply that the selective reflection of the red light should be observed at lower temperatures. Indeed, the red color arises in the images of the SmC$_A$* phase below ca. 280 K and remains in the SmC$_A$* glass; see the representative images registered in 260 K and 173 K in Figure 3. The numerical analysis (Figure 4) shows that the contribution of the red component increases on cooling from 280 K to 240 K. In comparison, below 240 K, all components are constant, which is attributed to the glass transition. It indicates that the helix pitch does not change noticeably with temperature in the vitrified state. On subsequent heating, an additional peak in the luminance is observed at 270-280 K, which is explained by nucleation. Due to the presence of the aligning layer, the sample returns to the previous alignment at ~300 K, corresponding to the crystal phase's melting temperature [10].

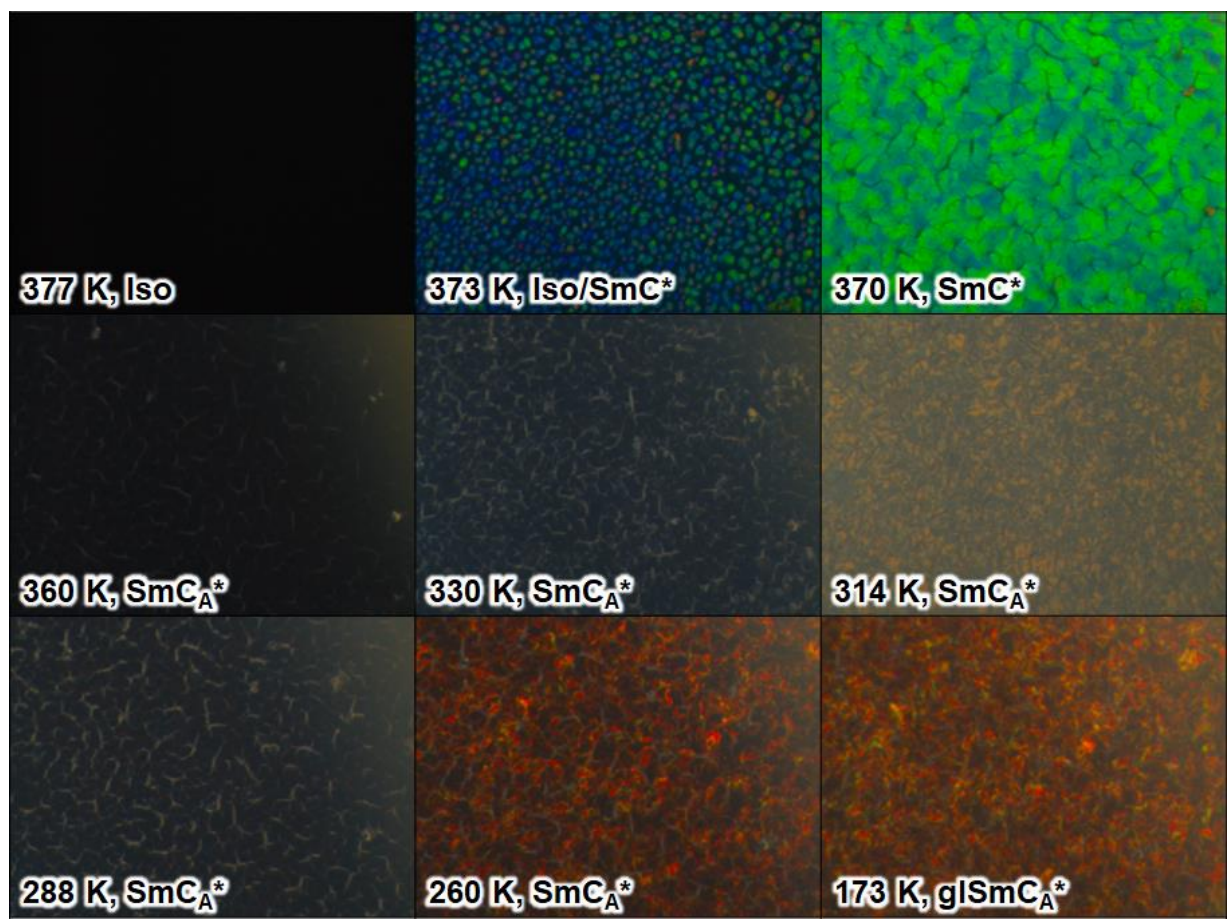

Figure 3. POM textures of 3F5HPhF6 registered on cooling at 10 K/min in the reflection mode. Each image shows an area of 622 × 466 μm$^2$.

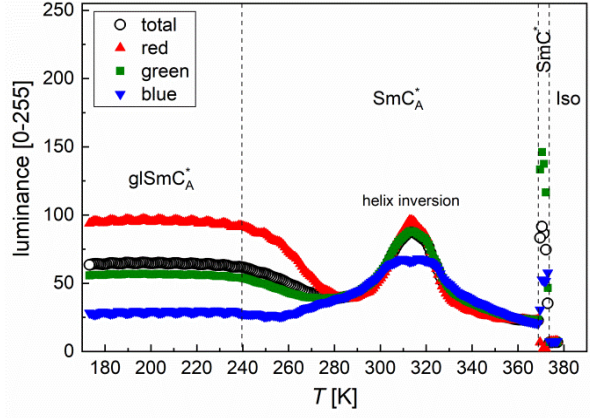

Figure 4. Average weighted luminance and separate contributions of the red, green, and blue components of the POM textures of 3F5HPhF6 obtained on cooling at 10 K/min in the reflection mode.

*3.3. X-ray diffraction and molecular modeling*

The XRD patterns of 3F5HPhF6 and the empty sample holder are shown in Figure 5. The sharp peak at $2\theta \approx 2.5°$ and broad maxima at $2\theta \approx 4.2°$ and $6.7°$ originate from the sample holder. The sharp peak at $2\theta \approx 2.9°$ and its higher harmonics, up to the 4$^{th}$ one, originate from the quasi-long range positional order in the SmC$_A$* phase, i.e., the smectic layers [1,2]. The broad maximum at $2\theta \approx 19$-$20°$ is related to the short-range positional order within the smectic layers [44]. In the heating run, numerous sharp peaks at 298 K indicate partial cold crystallization. The appearance of very small crystallites at lower temperatures, implied by the POM observations, is not visible in the XRD patterns. The Bragg equation [45] relates the positions $\theta_l$ of the (00$l$) peaks, where $l$ = 1, 2, 3, 4, to the smectic layer spacing $d$:

$$\theta_l = \theta_0 + \arcsin\left(\frac{l\lambda}{2d}\right), \tag{1}$$

where $\theta_0$ is the systematic shift in the peak positions and $\lambda$ = 1.540562 Å (CuKα1 characteristic wavelength [31]). The layer spacing was determined by fitting Equation (1) to the experimental peak positions. In temperatures 288 K and 298 K, only (001) and (003) peaks were visible, thus, the $\theta_0$ shift was taken from 278 K. The obtained results (Figure 6a) show that the layer spacing has a local maximum at 238 K, interpreted as a sign of the glass transition. In the glassy SmC$_A$* phase, the layer spacing decreases on cooling, with a relative decrease of 2.0% from 238 K to 18 K.

The short-range order in the smectic layers is described by the correlation length $\xi$ and the average distance $w$ between molecules. After recalculation of the $2\theta$ angle into the scattering vector $q = 4\pi \sin\theta/\lambda$, the wide maximum at $2\theta \approx 19$-$20°$ ($q_0 \approx 1.3$-$1.4$) is described by the Lorentz peak function [44]:

$$I(q) = \frac{A}{1+\xi^2(q-q_0)^2} + Bq + C, \tag{2}$$

where $A$ is peak height and $B$, $C$ are parameters of the linear background. The average distance, obtained as $w = 2\pi/q_0$, decreases on cooling, with a relative decrease of 5.9% from 298 K to 18 K (Figure 6b). The glass transition temperature cannot be inferred from the $w(T)$ plot, as neither extrema nor discontinuities are observed at $T_g$. The correlation length increases when cooling down to 218 K and shows a decreasing trend when cooling below 218 K (Figure 6c). The slight decrease of $\xi$ in the glassy SmC$_A$* state is caused by a simultaneous decrease in $w$. In the plot of the $\xi/w$ ratio, one can see that above 218 K, the short-range positional order increases on cooling, while it practically does not change below 218 K (Figure 6d). Both $d$ below $T_g$ and $w$ in the whole investigated range can be fitted empirically with an exponential function $y(T) = y_0 + a \exp(T/T_0)$ and their values extrapolated to 0 K ($y_0 + a$) are equal to $d_0 = 29.60(2)$ Å and $w_0 = 4.437(3)$ Å. The $\xi$ and $\xi/w$ values can be fitted with two linear functions, with the intersection points at 228(4) K and 224(3) K, respectively. The slope of $\xi/w$ below $T_g$ was fixed to zero. The average value of the $\xi/w$ in the glassy SmC$_A$* phase is 1.261(2) and corresponds to the nearest-neighbor correlations only.

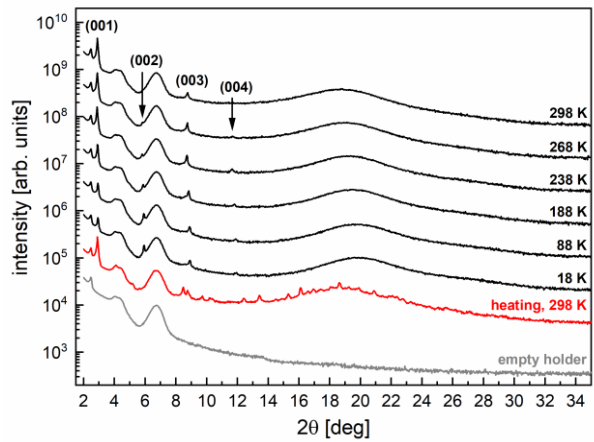

Figure 5. Selected XRD patterns of 3F5HPhF6 collected on cooling from 298 K to 18 K and after gradual heating to 298 K. The bottom pattern was collected at room temperature for an empty holder. The sharp peaks arising from the smectic layer order are indicated.

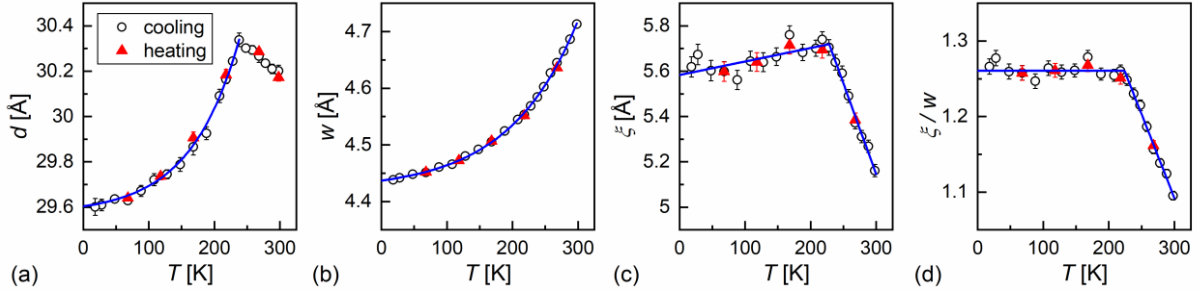

Figure 6. Smectic layer spacing (a), average intermolecular distance within the smectic layers (b), correlation length of the short-range order (c), and the ratio of the correlation length and intermolecular distance (d) as a function of temperature in the supercooled and vitrified SmC$_A$* phase of 3F5HPhF6. The legend in (a) is common for all panels.

The integrated intensity of the (001) peak from the smectic layers decreases on cooling (Figure 7a). The decrease in intensity accelerates below 250 K, which is around the glass transition region, and slows down below 88 K. The intensity of the (002) peak increases on cooling down to 68 K and slightly increases on further cooling. The intensity of the (003) peak has a broad maximum centered at 218 K, while the intensity of the (004) peak is approximately constant. The (001) peak is significantly stronger than the higher order peaks, caused by the Lorentz-polarization factors $Lp$, strengthening the intensities at low 2θ angles. The united form of the $Lp$ factors for the integrated intensities from the polycrystalline samples is [46-48]:

$$Lp = \frac{1+\cos^2(2\theta)}{\sin^2\theta \cos\theta}. \qquad (3)$$

The $Lp$-corrected intensities are calculated as $I_{corr} = I/(Lp)$. The relative $Lp$-corrected intensities of the (00$l$) peaks depend on the distribution of the electron density $\rho(z)$ along the smectic layer normal [1-3]. The molecules in the smectic phases can rotate around their short axes [49,50]. Thus, it can be assumed that the $\rho(z)$ distribution is centrosymmetric in relation to the middle of the smectic layer [51,52]:

$$\rho(z) = \rho_0 + \sum_{l=1}^{4} F_{00l} \cos(2\pi lz/d), \qquad (4)$$

where $F_{00l}$ is the structure factor. Generally, structure factors are complex numbers and the $\rho(z)$ function is described by complex $\exp(2\pi ilz/d)$ instead of cosines. The absolute $F_{00l}$ values are determined as $|F_{00l}| = \sqrt{I_{corr}}$ (Figure 7b), while their phases remain unknown. However, for the centrosymmetric $\rho(z)$ distribution, the simpler (4) formula can be applied, where $F_{00l}$ are real numbers, and the phase problem is reduced to the choice of either (+) or (−) sign [3,45,51]. With four diffraction peaks, it leads to $2^4 = 16$ combinations. Two sets will all opposite signs, for example (+,+,+,+) and (−,−,−,−), describe the same $\rho(z)$ distribution, only shifted by $d/2$ [51]. Since $\rho(z)$ has a period of $d$, it reduces the number of combinations to 8. Herein, we assume that a single smectic layer is located between $z = 0$ and 1, thus, the minimum of $\rho(z)$ is expected at $z = 0$. In this case, the $F_{001}$ factor should be negative (−).

In the SmC$_A$* phase in 298 K, there are two peaks, (001) and (003). The choice of the (−) sign of $F_{003}$ retains the minimum of $\rho(z)$ at $z = 0$, while the choice of the (+) sign leads to a local maximum of $\rho(z)$ at $z = 0$, which is against our earlier assumption. The signs of $F_{002}$ and $F_{004}$ are less apparent. The analysis for the results at lower temperatures shows that only the (−,+,−,+) combination has to be excluded because it gives the minima in $\rho(z)$ close to the middle of the smectic layer, while the (−,−,−,−), (−,+,−,−), and (−,−,−,+) combinations describe $\rho(z)$ with minima at $z = 0$ (Figure 8).

The obtained $|F_{00l}|$ values indicate that the electron density distribution along the layer normal evolves also below the glass transition temperature (Figure 7b). Above 168 K, the $|F_{001}|$ factor is dominant and below 168 K, it decreases and becomes practically equal to $|F_{003}|$. The $|F_{002}|$ and $|F_{004}|$ are smaller than $|F_{001}|$ and $|F_{003}|$ in the whole temperature range. Down to 148 K, $|F_{002}|$ is smaller than $|F_{004}|$, and below 148 K, this relationship reverses. Regardless of the choice of the $F_{002}$ and $F_{004}$ signs, $\rho(z)$ deviates strongly from the simple cosine function.

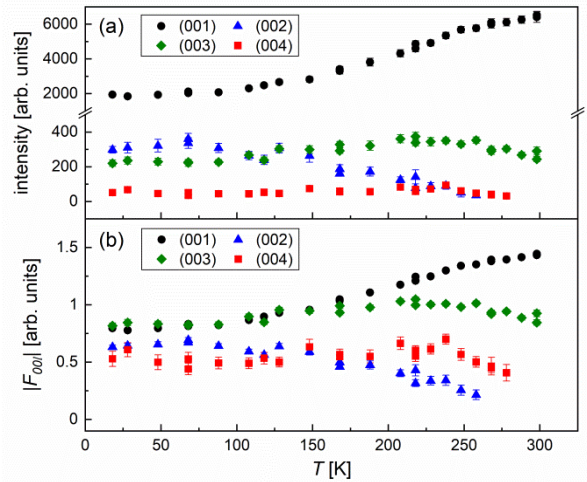

Figure 7. Integrated intensities of the (00$l$) peaks from the smectic layers of supercooled and vitrified 3F5HPhF6 (a) and the absolute values of the structure factors, calculated from these intensities after correction by the Lorentz-polarization factors (b).

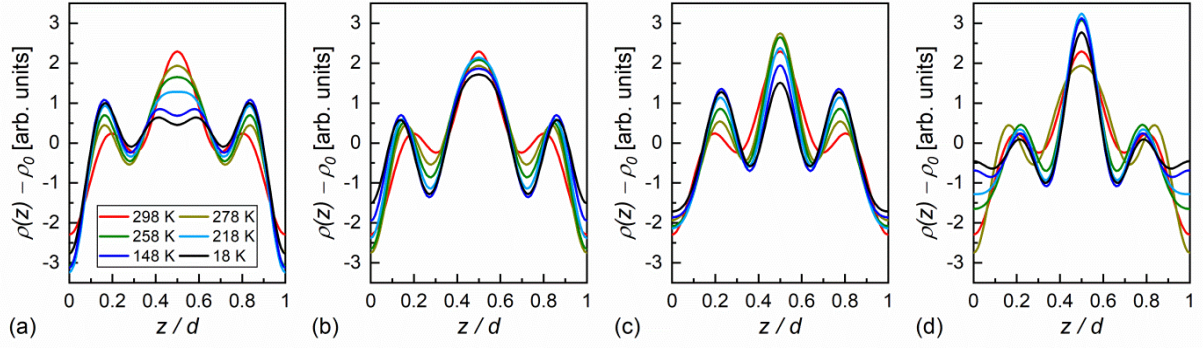

Figure 8. Electron density distribution along the smectic layer normal in 3F5HPhF6, calculated for various choices of the $F_{001}$, $F_{002}$, $F_{003}$, $F_{004}$ signs: (−,−,−,−) (a), (−,+,−,−) (b), (−,−,−,+) (c), and (−,+,−,+) (d). The legend in (a) is common for all panels.

The electron density distribution can be approximated based on the DFT calculations [53,54]. Figure 9 shows the 3F5HPhF6 molecular model. The single-crystal XRD results for the MHPOBC compound in a crystal phase [55,56] as well as the recently published semi-empirical and molecular dynamics calculations for the 3F5HPhF9 compound [13] show that the $C_nH_{2n}-$ chain forms an angle of ~90° with the molecular core. Such conformation was assumed for the 3F5HPhF6 molecule. The bending in the $-OCH_2C_3F_7$ part was included to match the experimental smectic layer spacing and the tilt angle of the molecular core equal to ca. 45° [10,14]. If one includes the 1.09 Å and 1.44 Å radii for the H and F terminal atoms [57], the layer spacing equals 29.06 Å.

The distribution of electron density around each atom was approximated by the cusp-like function $Z^2 \exp(-2Z|z-z_0|)$, based on the one mentioned in [58], where $Z$ is the atomic number corrected by the Mulliken partial charge calculated by DFT for each atom and the pre-exponential parameter $Z^2$ provides the area of the cusp equal to $Z$. The average electron density, calculated separately for the $C_6H_{13}-$, $-COOPhPhCOOPh-$, $-OC_5H_{10}-$, and $-OCH_2C_3F_7$ parts, is the highest for the fluorinated part of the achiral chain and the aromatic core.

Figure 10 presents the centrosymmetric electron density distribution, corresponding to the average over two 3F5HPhF6 molecules rotated by 180° with respect to the middle of the smectic layer. Equation (4) was fitted to obtain a smoother distribution. The obtained result resembles the $\rho(z)$ distributions determined from the XRD patterns for the $F_{001}$, $F_{002}$, $F_{003}$, and $F_{004}$ signs (−,−,−,−) (Figure 8a) and (−,+,−,−) (Figure 8b). The variant with the (−,−,−,+) choice of signs can be excluded at this stage, as it indicates the lateral maxima in $\rho(z)$ too close to the middle of the smectic layer. The DFT results show that these maxima originate from the fluorinated end of the achiral chain.

At this point, only the (−,−,−,−) and (−,+,−,−) variants are left. The fitting result in Figure 10 corresponds to the (−,+,−,−) variant. However, the obtained $F_{002}$ value is close to zero. Therefore, this result cannot be taken for granted, especially since it was performed mainly as a smoothing procedure. Nevertheless, the (−,−,−,−) and (−,+,−,−) variants lead to similar conclusions (Figure 8a,b): as the temperature decreases, the lateral maxima in $\rho(z)$ shift towards the borders of the smectic layer and the central maximum lowers. The $(-,-,-,-)$ variant indicates additionally a slight split of the central maximum deep in the glassy state.

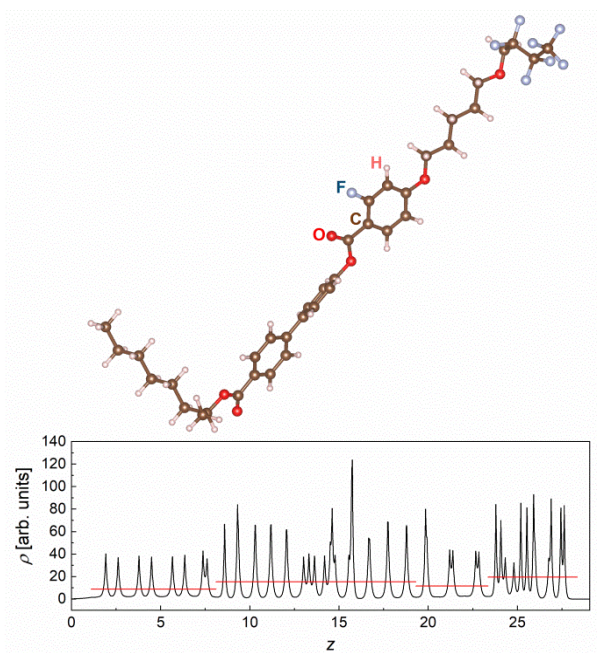

Figure 9. The model of the 3F5HPhF6 molecule optimized by the DFT method (B3LYP-D3(BJ) functional, def2TZVPP basis set) and the corresponding electron density profile projected on the smectic layer normal, oriented in the horizontal direction. The red lines indicate the average electron density for various molecular parts.

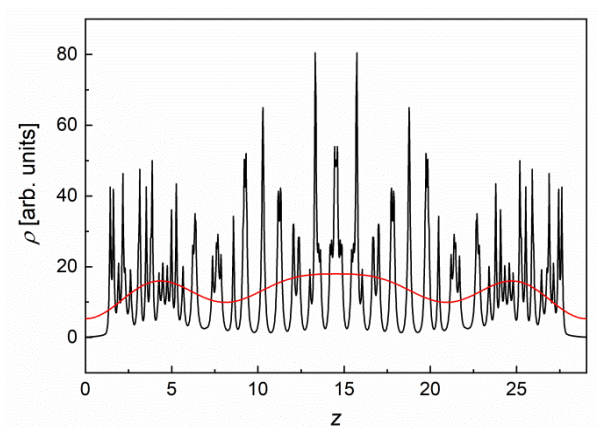

Figure 10. Centrosymmetric electron density distribution of 3F5HPhF6 based on the DFT calculations (B3LYP-D3(BJ), def2TZVPP) along the smectic layer normal. The red line is the fitting result of Equation (4).

**4. Summary and conclusions**

The structural studies of the vitrified SmC$_A$* phase in the 3F5HPhF6 compound were carried out using the XRD method and DFT calculations. The smectic layer spacing shows a local maximum in the glass transition region. In contrast, the average distance between molecules within layers decreases on cooling without any anomalies around the glass transition temperature. The correlation length of the short-range positional order within layers remains constant in the vitrified state with respect to the average intermolecular distance.

The electron density profile is assumed to be centrosymmetric because the molecules can rotate around their short axes. The intensities of diffraction peaks from the smectic layer order show the comparable contribution of the $\cos(2\pi z/d)$ and $\cos(6\pi z/d)$ components to the electron density profile along the smectic layer normal, which results in three local maxima, one in the center and two close to the borders of the smectic layer. The central maximum corresponds to the fluorinated molecular core and the lateral maxima correspond to the fluorinated end of the achiral chain. The $F_{001}$, $F_{003}$, and $F_{004}$ structure factors are determined to be negative because only this combination provides the electron density distribution in agreement with the DFT calculations. The sign of the $F_{002}$ factor remains ambiguous. Despite this, both the (−,−,−,−) and (−,+,−,−) variants indicate the shift of the fluorinated chain ends towards the borders of the smectic layers with decreasing temperature.

The microscopic observations of the selective reflection show that the helix pitch in the SmC$_A$* phase is practically constant below the glass transition temperature. It can be assumed that any undetected, small changes in the pitch might be the effect of the shrinking in the smectic layer spacing, as the helix axis is perpendicular to the smectic layers (along the smectic layer normal).

The presented results show that the glass transition affects the properties of the SmC$_A$* phase to various extents. Also, the determined glass transition temperature may differ depending on the investigated property. The dynamic $T_g$ corresponding to the α-relaxation time equal to 100 s is equal to 229-230 K [15]; the calorimetric $T_g$ determined from the middle in the step of the heat capacity is equal to 227-231 K; and the temperature evolution of the intra-layer short-range order indicates $T_g$ = 224-228 K. Meanwhile, the smectic layer spacing has a local maximum at $T_g$ = 238 K and the observations of selective reflection, as well as the POM measurements, indicate $T_g$ = 240-245 K. It suggests that the structure along the smectic layer normal – layer spacing at the molecular scale and helix pitch at the mesoscopic scale – are affected by the glass transition at higher temperatures, at least for 3F5HPhF6.

**Acknowledgment:** The study was carried out using research infrastructure funded by the European Union in the framework of the Smart Growth Operational Programme, Measure 4.2; Grant No. POIR.04.02.00-00-D001/20, "ATOMIN 2.0 – Center for materials research on ATOMic scale for the INnovative economy". We gratefully acknowledge Polish high-performance computing infrastructure PLGrid (HPC Center: ACK Cyfronet AGH) for providing computer facilities and support within computational grant no. PLG/2024/017946.

**Authors' contributions:**

A. Deptuch – conceptualization, investigation, formal analysis, writing – original draft

M. Kozieł – investigation, writing – review and editing

M. Piwowarczyk – investigation, writing – review and editing

M. Urbańska – resources, writing – review and editing

E. Juszyńska-Gałązka – investigation, writing – review and editing

# Low-temperature structural study of smectic $C_A$* glass by X-ray diffraction

Aleksandra Deptuch[1,*], Marcin Kozieł[2], Marcin Piwowarczyk[1], Magdalena Urbańska[3], Ewa Juszyńska-Gałązka[1,4]

[1] Institute of Nuclear Physics Polish Academy of Sciences, Radzikowskiego 152, PL-31342 Kraków, Poland

[2] Faculty of Chemistry, Jagiellonian University, Gronostajowa 2, PL-30387, Kraków, Poland

[3] Institute of Chemistry, Military University of Technology, Kaliskiego 2, PL-00908 Warsaw, Poland

[4] Research Center for Thermal and Entropic Science, Graduate School of Science, Osaka University, 560-0043 Osaka, Japan

* corresponding author, aleksandra.deptuch@ifj.edu.pl

# Supplementary Materials

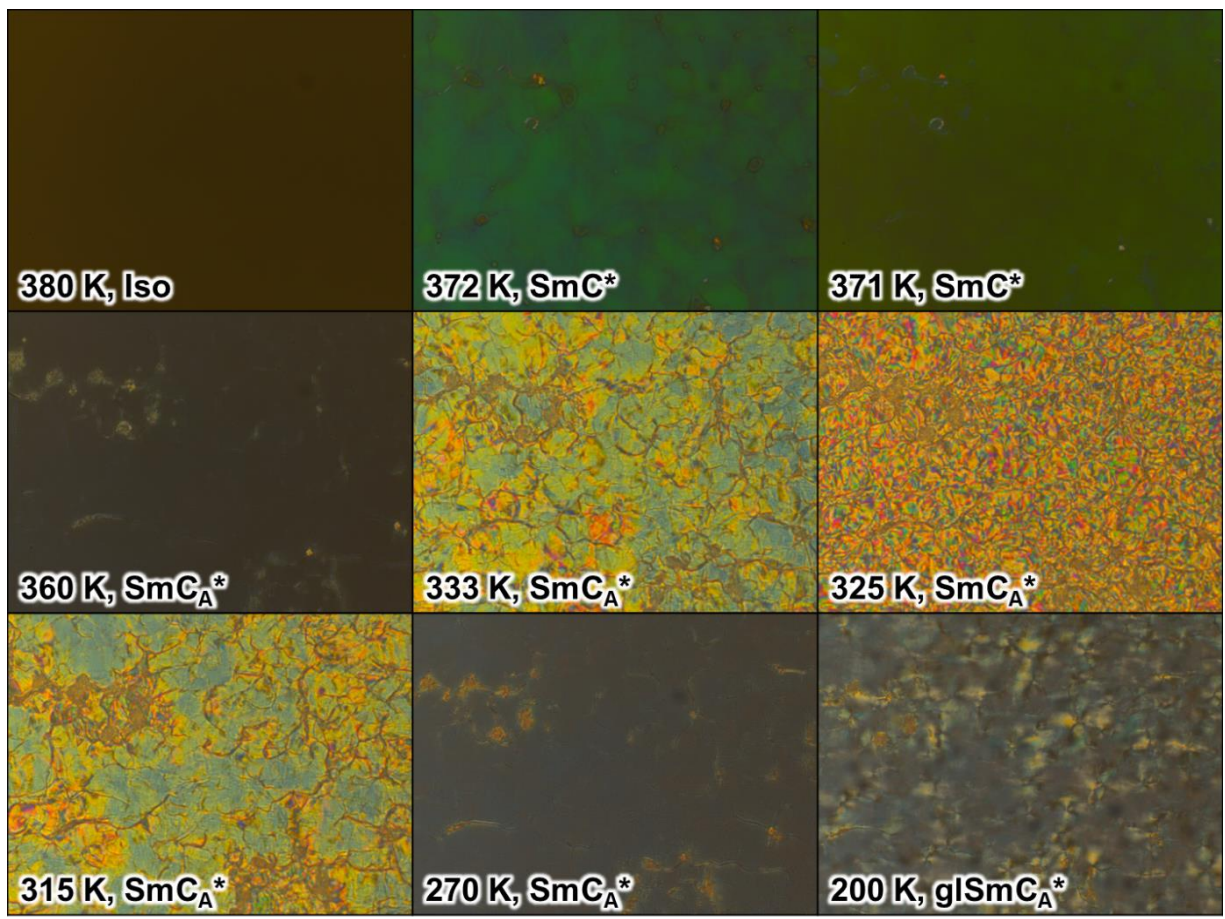

Figure S1. POM textures of 3F5HPhF6 registered on cooling at 10 K/min in the transmission mode. Each image shows an area of 1243 × 933 μm².

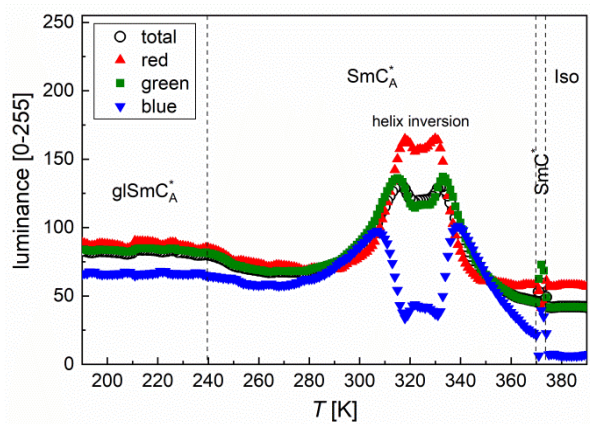

Figure S2. Average weighted luminance and separate contributions of the red, green, and blue components of the POM textures of 3F5HPhF6 obtained on cooling at 10 K/min in the transmission mode.

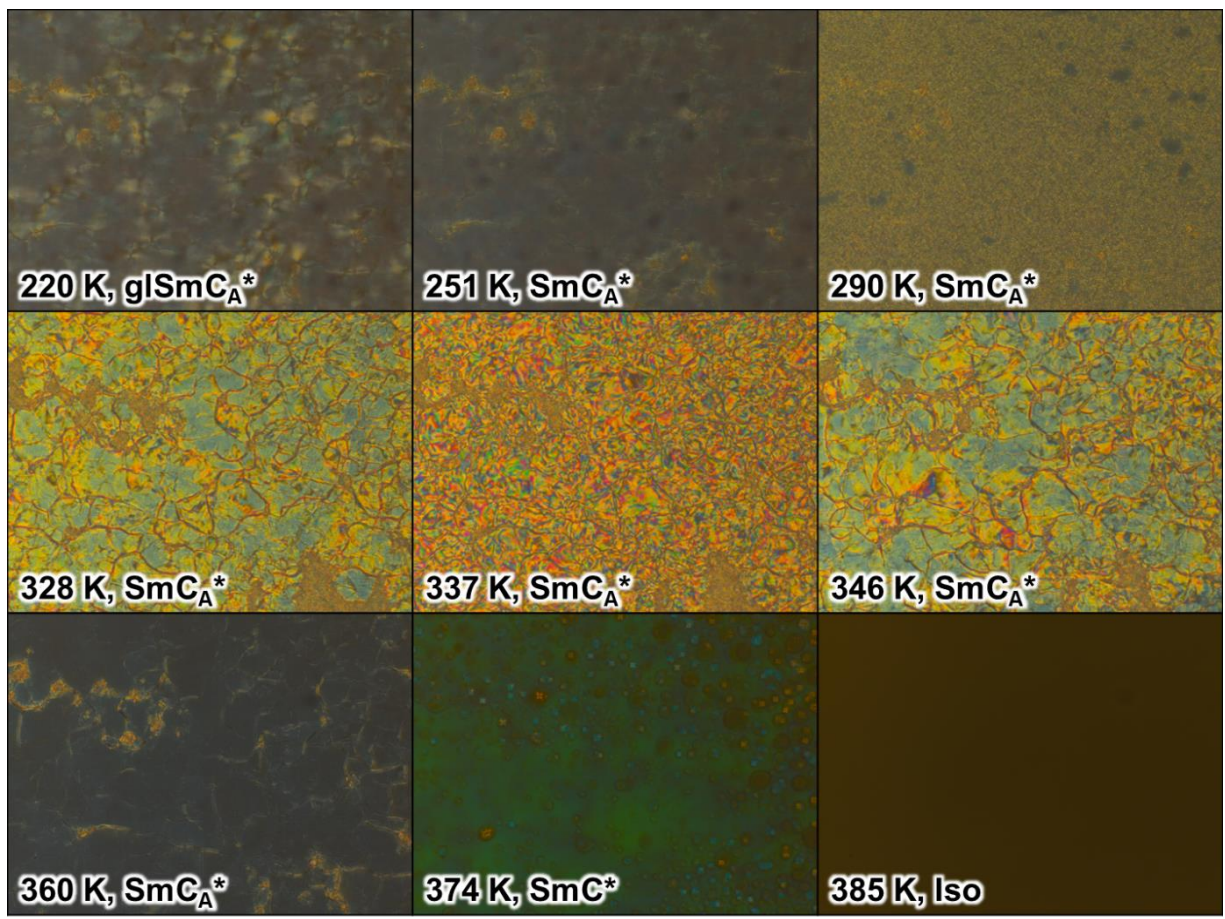

Figure S3. POM textures of 3F5HPhF6 registered on heating at 10 K/min in the transmission mode. Each image shows an area of 1243 × 933 μm$^2$.

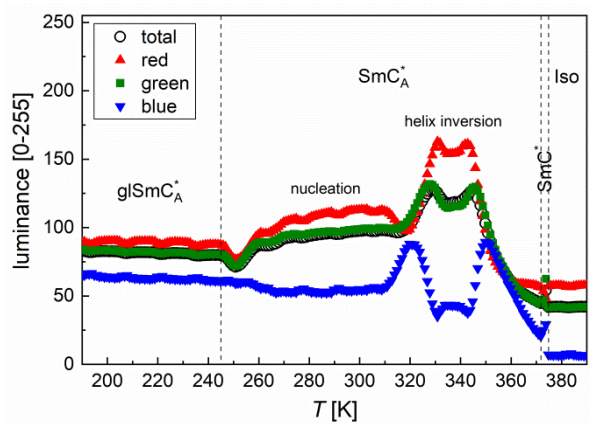

Figure S4. Average weighted luminance and separate contributions of the red, green, and blue components of the POM textures of 3F5HPhF6 obtained on heating at 10 K/min in the transmission mode.

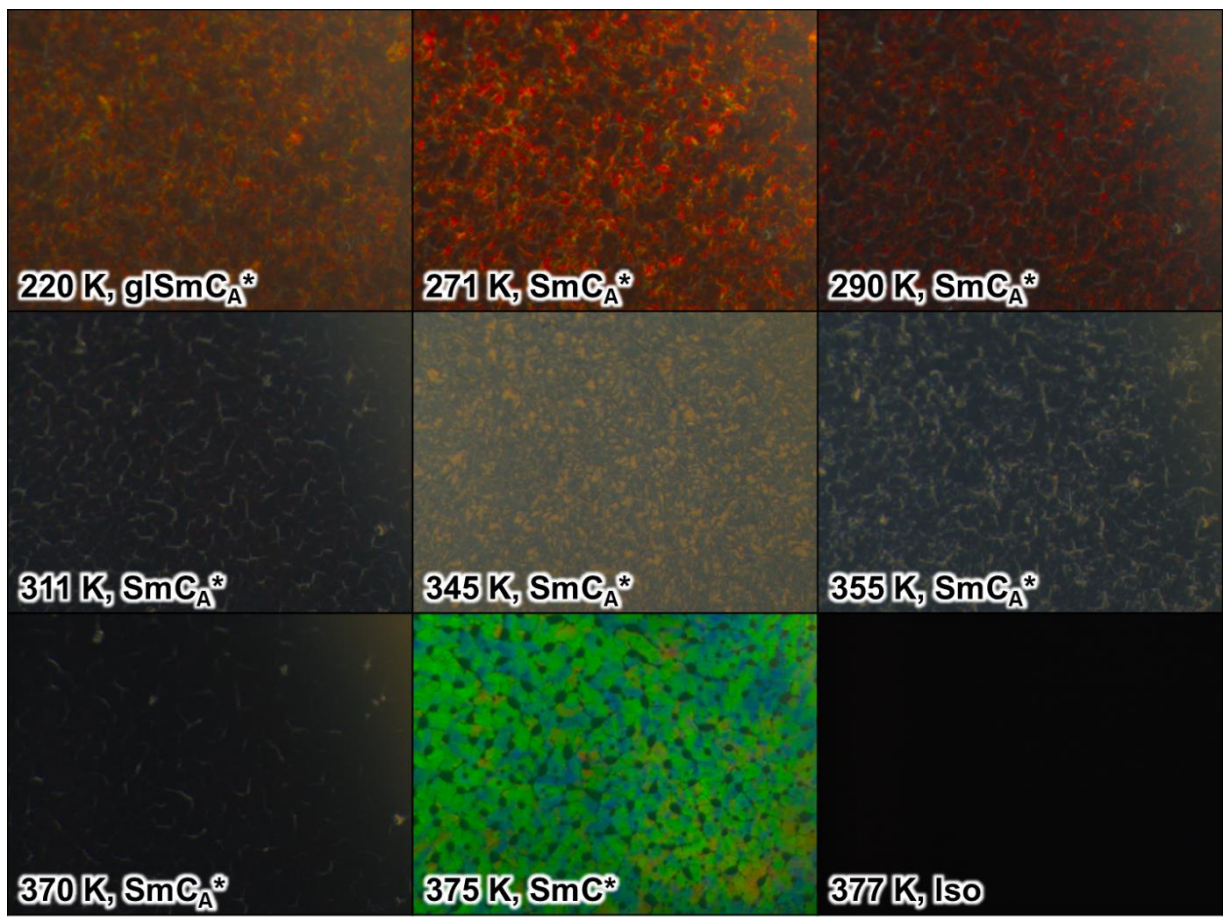

Figure S5. POM textures of 3F5HPhF6 registered on heating at 10 K/min in the reflection mode. Each image shows an area of 622 × 466 μm$^2$.

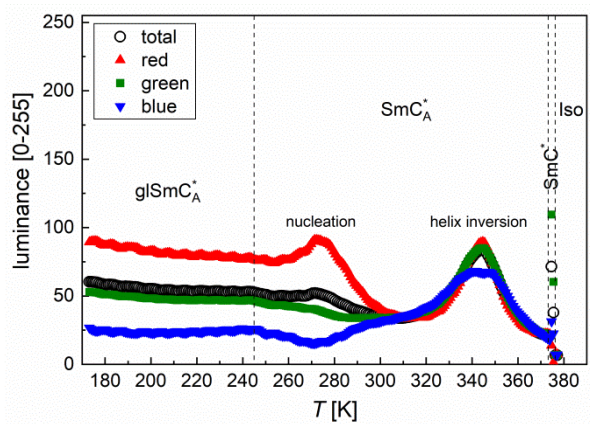

Figure S6. Average weighted luminance and separate contributions of the red, green, and blue components of the POM textures of 3F5HPhF6 obtained on heating at 10 K/min in the reflection mode.

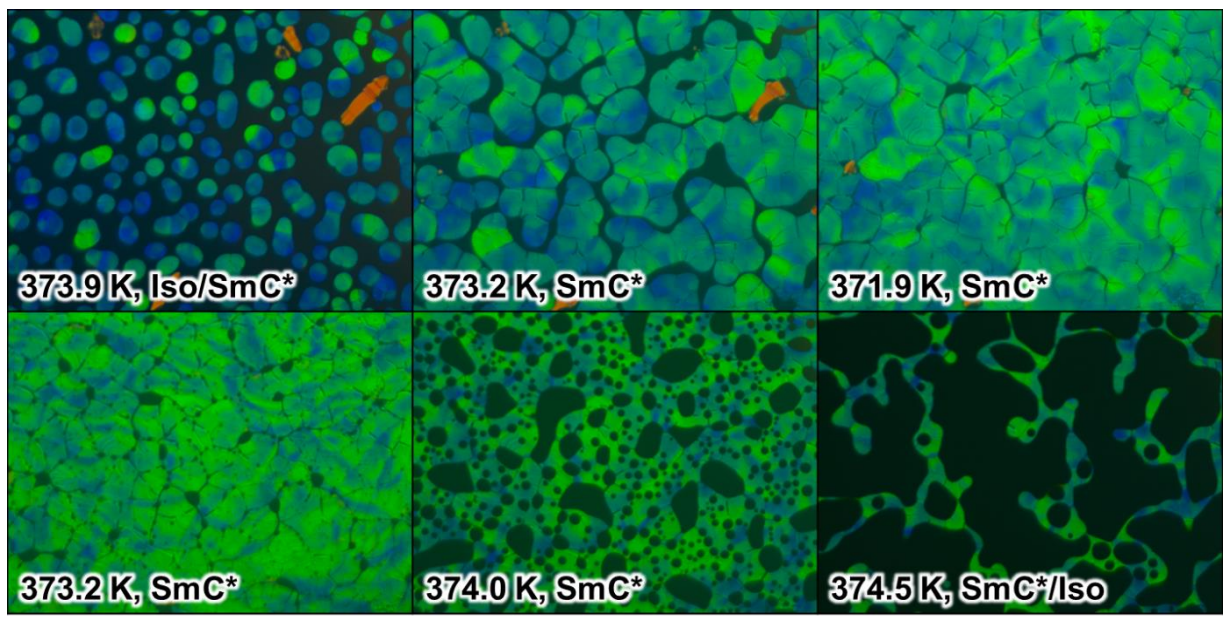

Figure S7. POM textures of 3F5HPhF6 registered on cooling (upper row) and heating (bottom row) at 1 K/min in the reflection mode. Each image shows an area of 622 × 466 μm$^2$.

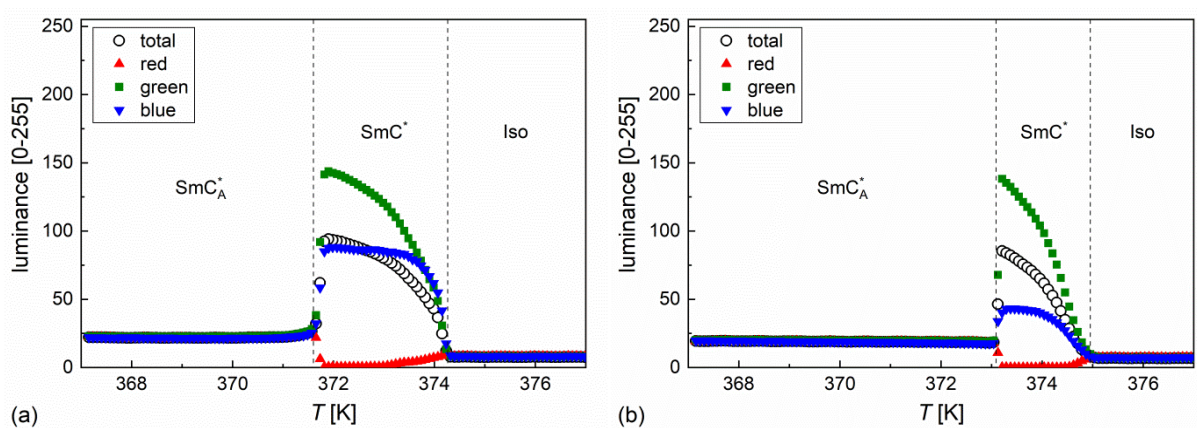

Figure S8. Average weighted luminance and separate contributions of the red, green, and blue components of the POM textures of 3F5HPhF6 obtained on cooling (a) and heating (b) at 1 K/min in the reflection mode.